\documentclass[english,aps,prl,twocolumn]{revtex4-1}
\pdfoutput=1
\usepackage[T1]{fontenc}
\usepackage[latin9]{inputenc}
\usepackage{amsmath}
\usepackage{amssymb}

\makeatletter
\usepackage{tikz}
\usetikzlibrary{hobby}
\usetikzlibrary{cd}

\makeatother

\usepackage{babel}
\begin{document}
\title{A geometric formulation of Schaefer's theory of Cosserat solids}
\author{Bal\'azs N\'emeth}
\author{Ronojoy Adhikari}
\affiliation{Department of Applied Mathematics and Theoretical Physics, Centre
for Mathematical Sciences, University of Cambridge, Wilberforce Road,
Cambridge CB3 0WA, United Kingdom}
\begin{abstract}
The Cosserat solid is a theoretical model of a continuum whose elementary
constituents are notional rigid bodies, having both positional and
orientational degrees of freedom. Cosserat elasticity has had a revival
of interest stemming from applications in soft robotics and active
soft matter. Here we present a formulation of the mechanics of a Cosserat
solid in the language of modern differential geometry and exterior
calculus, motivated by Schaefer's ``motor field'' theory. The solid
is modelled as a principal fibre bundle, with the base labelling the
positions of the constituents and the fibre accommodating their orientations.
Configurations of the solid are related by translations and rotations
of each constituent. This essential kinematic property is described
in a coordinate-independent manner using a bundle map. Configurations
are equivalent if this bundle map is a global isometry of Euclidean
space. Inequivalent configurations, representing deformations of the
solid, are characterised by the local structure of the bundle map.
Using Cartan's magic formula we show that the strain associated with
infinitesimal deformations is the Lie derivative of a connection one-form
on the bundle. The classical infinitesimal strain is thus revealed
to be a Lie algebra-valued one-form. Extending Schaefer's theory,
we derive the strain associated with finite deformations by integrating
the infinitesimal strain along a prescribed path. This is path independent
when the curvature of the connection one-form is zero. Path dependence
signals the presence of topological defects and the non-zero curvature
is then recognised as the density of topological defects. Mechanical
stresses are defined by a virtual work principle in which the Lie
algebra-valued strain one-form is paired with a dual Lie algebra-valued
stress two-form to yield a scalar work volume form. The d'Alembert
principle for the work form provides the balance laws, which we obtain
in the limit of vanishing inertia. The work form is shown to be integrable
for a hyperelastic Cosserat solid and the breakdown of integrability,
relevant to active oriented solids, is briefly examined. Our work
elucidates the geometric structure of Schaefer's theory of the Cosserat
solid, aids in constitutive modelling of active oriented materials,
and suggests structure-preserving integration schemes for numerical
simulation.
\end{abstract}
\maketitle

\section{Introduction\label{sec:Introduction}}

The Cosserat solid is a model of a continuum in which the material
constituent has both positional and orientational degrees of freedom.
It was presented by E. and F. Cosserat in their 1909 opus \textit{\textquotedbl Th\'eorie
des Corps deformables\textquotedbl}. The Cosserats' approach to
elasticity lay neglected for half a century following publication
before it was revisited in the middle of the twentieth century in
the context of the microstructured materials \citet{Ericksen1957},
\citet{Toupin1962}, \citet{Mindlin1964}, \citet{Schaefer1967} and
\citet{Kroner1968}. We refer the reader to the lucid historical reviews
of Schaefer \citep{SchaeferReview} and Eringen \citep{Eringen1999}. 

Cosserat elasticity makes an appearance in effective descriptions
of rods and shells undergoing large deformations \citep{Reissner1981,Simo1988,Simo1985}.
Orientational degrees of freedom emerge when a dimensional reduction
is employed to represent the deformation of the body as a combination
of translations and rotations of its rigid cross-sections. Cosserat
rod and shell theories have been greatly successful in modelling deformations
of soft and slender structures including biological filaments \citep{Floyd2022},
membranes or active metabeams \citep{Chen2021}. The underlying theoretical
principles of the Cosserats has been generalised in many directions
\citep{Capriz1989}. Recent examples include extensions of general
relativity \citep{Hehl1995}, fracton gauge theories \citep{Gromov2020}
and topological defects in complex media \citep{Bohmer2021,Randono2011}.
Cosserat elasticity was an inspiration for Elie Cartan in his approach
to differential geometry \citep{Hehl2007}. 

Despite its theoretical interest, the experimental validity of the
Cosserat solid remained in question for many years. Recent advances
in 3D printing techniques has made it possible to construct metamaterials
whose material response is consistent with Cosserat elasticity \citep{Rueger2018,Reasa2020,Frenzel2017,Chen2020}.
Active soft matter provides many experimental systems which contain
both translational and rotational degrees of freedom and Cosserat
elasticity, with suitably chosen constitutive assumptions \citep{Surowka2022,Bolitho2022}
may be used to model them. The additional rotational degrees of freedom
in Cosserat theory allow for the long-wavelength description of chiral
materials, whether active or passive. Such materials cannot, \emph{a
priori,} be described by Cauchy elasticity in which only translational
degrees of freedom are retained. 

Given the theoretical and experimental importance of the Cosserat
solid, it is surprising that the mathematical underpinnings of the
theory have remained veiled. While the Cosserats used a most cumbersome
notation that is hard to fathom for the modern reader, the majority
of the following treatments relied on local coordinates and Cartesian
tensors. This has led to a profusion of notation and confusion even
over the fundamental strain measures \citep{Pietraszkiewicz2009}.
The pioneering work of Schaefer \citep{Schaefer1967} stands out as
a notable exception in being one of the earliest treatments that attempted
to extract the geometric content of the Cosserat theory and express
it in the language of differential forms. Focussing on infinitesimal
deformations, Schaefer absorbed the infinitesimal displacement and
rotation fields of a Cosserat solid into ``motor'' fields. The ``motors''
were taken to be elements of a six-dimensional vector space of infinitesimal
translations and rotations, and had been introduced by von Mises in
his treatment of rigid body mechanics \citep{Mises1924}. Strain was
defined as a differential one-form measuring the deviation of the
motor field from an infinitesimal rigid transformation, the latter
defining a parallelism and a covariant derivative. Schaefer recognized
the importance of duality between forces and velocities: he defined
stress as a differential two-form taking values in the dual space
of motors. This was motivated by the observation that forces should
generally be thought of as covectors taking values in the vector space
dual to velocities, with the duality pairing giving the rate of power
expended by the force \citep{Kanso2007,Frankel2011,Rashad2023}. For
the Cosserat solid, the pairing of stress with a velocity field (represented
by a motor field) yields a scalar-valued two-form that can be integrated
on the boundary of the solid giving the rate of work done by stresses.
Balance laws and and topological defects in Cosserat media were discussed
within the framework provided by the motor calculus, i.e the combination
of the algebra of motors and the calculus of differential forms.

Schaefer's approach, despite its originality, economy and clarity
has not been widely adopted, probably owing to the fact that motor
calculus is much less known and appreciated in the continuum mechanics
and soft matter community than tensor calculus. Further, his derivations
are often based on heuristic arguments and analogies, rather than
established mathematical constructions. To remedy we present Schaefer's
theory of the Cosserat solid in the language of modern differential
geometry. The key mathematical concepts that appear are Lie groups
and their algebras, principal bundles, and differential calculus that
results from combining these structures. From this perspective, a
nonlinear theory of the Cosserat solid, absent in Schaefer's theory,
becomes available and linearisation in Schaefer's theory is elucidated. 

We provide a brief survey of our formulation before presenting the
details below. Fibre bundles are manifolds which have a local product
structure: any point of the bundle has a neighbourhood which looks
like a product manifold of the form $U\times F$, with $U$ being
an open subset of the base $B$ of the bundle and another manifold
$F$, called a typical fibre of the bundle \citep{Frankel2011}. Two
especially important classes of fibre bundles are vector bundles (when
$F$ is a vector space) and principal bundles (when $F$ is a Lie
group). In continuum mechanics, fibre bundles can naturally model
media whose material particles possess a complex inner structure:
the base manifold $B$ represents the material particles and the fibre
$E$ is the collection of all possible configurations of the microstructure
\citep{Segev1994}. They are also the underlying mathematical model
in most field theories of soft condensed matter physics: the base
$B$ is the ambient space while the fibre $F$ is an order parameter
manifold corresponding to some broken symmetry \citep{Chaikin1995}.
In the case of a Cosserat solid, the fibre bundle modelling the body
is a principal bundle $P$ \citep{Badur1998,Binz1998,Epstein1998,Delphenich2012}
(as $F$ can be identified with the orthogonal group) and therefore
has a much richer structure than a general fibre bundle. We will show
that Schaefer's space of motor fields is in fact a vector bundle associated
to $P$, with the typical vector space being the Lie algebra of the
Euclidean group, thus giving a precise meaning to motors. Invariance
under rigid transformations will lead us to investigate the Maurer-Cartan
form on the Euclidean group and view it as a Cartan connection $\omega$
on $P$ \citep{Sharpe1997}. It will turn out that strain is the result
of any changes in this connection form along a deformation -- in
particular, Schaefer's covariant derivative on the bundle of motor
fields is induced by $\omega$. Topological defects will be related
to the curvature of the connection. Stress as a differential $2$-form
taking values in the dual bundle of motors will be introduced via
a duality argument outlined above (for similar formulations of classical
elasticity in terms of vector bundle valued differential forms, see
\citep{Kanso2007,Rashad2023}). Exploiting this duality, equations
of motion will be derived from a virtual work principle. Since in
the future we intend to apply our theory to active solids which are
generally overdamped \citep{Ferrante2013,Maitra2019,Xu2022}, we will
not consider inertial forces.

A remarkable property of our formulation is that it only relies on
the existence of a connection and does not directly use the metric
of Euclidean space. This is a manifestation of the hierarchy of geometric
structures: a Riemannian metric in differential geometry is a high-level
structure because it automatically gives rise to a unique connection
through the Levi-Civita construction as well as providing an isomorphism
between tangent and cotangent spaces and a distinguished volume form
for orientable manifolds. It is interesting that a more low-level
structure, namely a connection (which roughly speaking provides a
way to parallel transport vectors along curves) is sufficient to describe
deformations in these complex media.

The article is organized as follows. In Section II, the necessary
preliminaries are given about the representation of Cosserat solids
by principal bundles. In Section III the theory of strain is outlined
while in Section IV compatibility conditions are discussed. In Section
V stress is introduced and balance laws are derived from a virtual
work principle. Finally, in Section VI constitutive laws are touched
upon and conclusions are drawn in Section VII. We point the reader
to the comprehensive textbook \citep{Frankel2011}, which has an excellent
section on geometric continuum mechanics in terms of bundle-valued
differential forms; and to \citep{Flanders1989} which provides a
brilliant introduction to exterior calculus.

\section{Preliminaries\label{sec:Preliminaries}}

\begin{figure}
\begin{tikzpicture}
 	\draw[rounded corners] (0,0) .. controls (-.3,.4) and (-.2,.7) .. (0,2) .. controls (1.3,2.3) and (2.2, 2.2) .. (3,2) .. controls (3.3,1.7) and (3.1,0.4) .. (3,0) -- cycle;
	\node at (.2,.2) {$\mathcal{B}$};
	\filldraw (2,1) circle[radius=1.5pt] node[below] {$X$};
	\draw[->,dashed] (2,1) .. controls (3.5,2.3) and (4,2) .. node[below] {$\kappa$} (4.8,1.8);
	\filldraw (4.8,1.8) circle[radius=1.5pt] node[right] {$\kappa(X)$};
	\draw[dotted] (4,-1) rectangle (8,3.5) node[below left] {$\mathbb{E}^3$};
	\draw[rounded corners] (4.4,0) .. controls (4.2,1) and (4.1,1.3) .. (4.4,2.4) .. controls (5.2,2.1) and (6.7, 2.7) .. (7.4,2.4) .. controls (8,1.8) and (6.5,1.9) .. (7.4,0) .. controls (6,-.4) and (5,-.7) .. (4.4,0);
	\node at (5,.2) {$\kappa(\mathcal{B})$};
\end{tikzpicture}

\caption{Configuration of a simple body. In general, $\mathcal{B}$ is assumed
to be just a smooth manifold that labels the material particles, but
in many applications, $\mathcal{B}$ is identified with a submanifold
of $\mathbb{E}^{3}$ with the aid of a reference configuration.}
\end{figure}
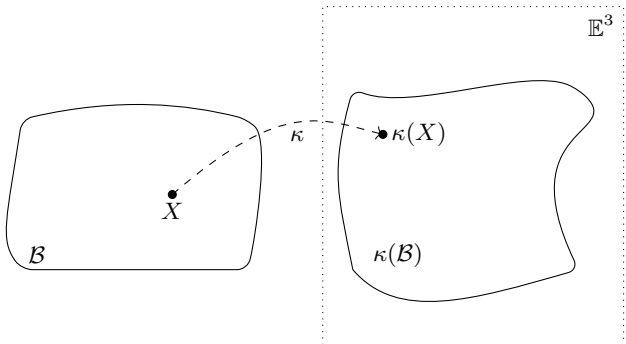
In continuum mechanics, a body is modelled as a smooth 3-manifold
$\mathcal{B}$ (often called the material manifold), and a configuration
of the body is an embedding $\kappa:\mathcal{B}\to\mathbb{E}^{3}$
into the ambient physical space $\mathbb{E}^{3}$, which is an affine
space with underlying translational vector space $\mathcal{E}$ \citep{Noll1974}.
Even though $\mathcal{E}$ is usually equipped with an inner product,
we will not make direct use of it in the sequel, because, as we will
demonstrate, strain and stress in a Cosserat solid are related to
the action of the Euclidean group, not deformations in the metric
structure. It is common practice to single out a reference configuration
$\kappa_{0}:\mathcal{B}\to\mathbb{E}^{3}$ and label material points
$X\in\mathcal{B}$ with their occupied position $x=\kappa_{0}(X)\in\mathbb{E}^{3}$
in the reference configuration, thereby identifying $\mathcal{B}$
with $\kappa_{0}(\mathcal{B})\subset\mathbb{E}^{3}$ \citep{Marsden1994}.

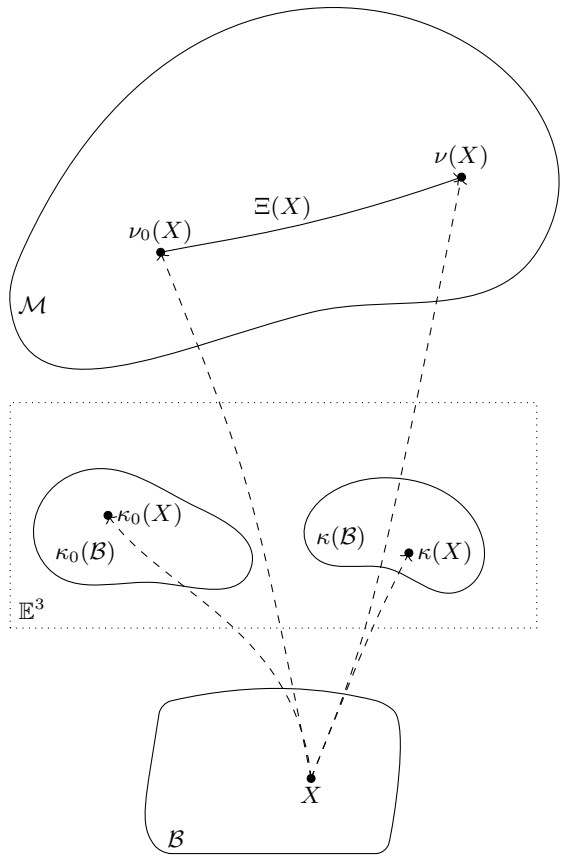
\begin{figure}
\begin{tikzpicture}
	\draw[rounded corners] (0,0) .. controls (-.3,.4) and (-.2,.7) .. (0,2) .. controls (1.3,2.3) and (2.2, 2.2) .. (3,2) .. controls (3.3,1.7) and (3.1,0.4) .. (3,0) -- cycle;
	\node at (.2,.2) {$\mathcal{B}$};
	\filldraw (2,1) circle[radius=1.5pt] node[below] {$X$};

	\draw[dotted] (-2,3) node[above right] {$\mathbb{E}^3$} rectangle (5,6);
	\draw (-1.5,3.8) to[closed,curve through ={(-1,5.1) .. (0.5,4.6) .. (1.2,3.8)}] (0,3.6);
	\node at (-1,4) {$\kappa_0(\mathcal{B})$};
	\filldraw (-.7,4.5) circle[radius=1.5pt] node[right] {$\kappa_0(X)$};
	\draw[->,dashed] (2,1) .. controls (1.8,3) and (0,3.5) .. (-.7,4.5);
	\draw (2,4) to[closed,curve through={(3,5) .. (4,3.5)}] (3,3.8);
	\node at (2.4,4.2) {$\kappa(\mathcal{B})$};
	\filldraw (3.3,4) circle[radius=1.5pt] node[right] {$\kappa(X)$};
	\draw[->,dashed] (2,1) .. controls (2.8,3) .. (3.3,4);
	\draw (-2,7.3) to[closed,curve through={(-1.8,8.1) .. (5,8)}] (2,7.2);
	\node at (-1.7,7.3) {$\mathcal{M}$};
	\filldraw (0,8) circle[radius=1.5pt] node[above] {$\nu_0(X)$};
	\filldraw (4,9) circle[radius=1.5pt] node[above] {$\nu(X)$};
	\draw[->,dashed] (2,1) .. controls (1.2,5) .. (0,8);
	\draw[->,dashed] (2,1) .. controls (2.8,3) .. (4,9);
	\draw[->] (0,8) .. controls (1,8.2) and (2,8.3) .. node[above] {$\Xi(X)$} (4,9);
\end{tikzpicture}

\caption{Configuration of a body with microstructure in the ``traditional''
formalism \citep{Capriz1989}.}
\end{figure}
However, this model only captures the translational degrees of freedom
of the material particles. If the body has a more complex inner structure,
an additional map $\nu:\mathcal{B}\to\mathcal{M}$ on top of $\kappa$
is introduced to describe the configuration of the microstructure,
where $\mathcal{M}$ is the smooth manifold consisting of all possible
microstructure configurations \citep{Capriz1989}. Well-known examples
are polar liquid crystals with $\mathcal{M}=S^{2}$ (the two-sphere)
or nematics with $\mathcal{M}=\mathbb{RP}^{2}$ (the real projective
plane) \citep{Chaikin1995}. For Cosserat solids, $\mathcal{M}$ is
generally taken to be $\mathrm{SO(3)}$ (the Lie group of rotations
in $\mathbb{E}^{3}$). This captures the property that the material
points possess rotational degrees of freedom on top of translational
ones. To describe a deformation with respect to a reference configuration
$\kappa_{0},\nu_{0}$, one also needs to introduce a map which assigns
to each element $X\in\mathcal{B}$ a transformation $\Xi(X)$ of $\mathcal{M}$
that takes $\nu_{0}(X)$ to $\nu(X)$. (It is usually assumed that
a group $K$ of transformations acts transitively on $\mathcal{M}$,
so $\Xi(X)$ can be taken to be an element of $K$).

\begin{figure}
\begin{tikzpicture}
	\draw (-2,-2) to[closed,curve through={(-2.3,0.6) .. (-1.1,.8) .. (0,.4)}] (0,-1.8);
	\node at (-2.3,-1.3) {$\mathcal{B}$};
	\draw[dotted] (1,-2.7) node[above right] {$\mathbb{E}^3\cong G/H$} rectangle (4.5,2.2);
	\draw (1.3,-1.7) to[closed,curve through={(1.9,0) .. (2.8,1)}] (3.3,-1);
	\draw[->] (1.5,1.2,0) -- (1.5,1.2,1) node[below right] {$\mathbf{e}_a$};
	\draw[->] (1.5,1.2,0) -- (1.5,1.8,0);
	\draw[->] (1.5,1.2,0) -- (2.1,1.2,0);
	\node at (1.6,1.3) {$o$};
	\filldraw (-1,-1) circle[radius=1pt] node[below] {$X$};
	\filldraw (2.7,0.1) circle[radius=1pt] node[below] {$\kappa(X)$};
	\draw[->,dashed] (-1,-1) -- node[below] {$\kappa$} (2.7,0.1);

	\draw (-1,4) circle[radius=1.5];
	\node at (-2,3.2) {$H$};
	\draw[dotted] (-1,-1) -- (-2.431,3.55);
	\draw[dotted] (-1,-1) -- (0.431, 3.55);
	\draw[->] (-.8,4.2,0) -- (-.8,4.2,1);
	\draw[->] (-.8,4.2,0) -- (0,4.2,0);
	\draw[->] (-.8,4.2,0) -- (-.8,5,0);
	\filldraw (-.8,4.2) circle[radius=1pt] node[below right] {$p$};
	\draw[->, dashed] (-.8,4.2) -- node[left] {$\pi$} (-1,-1);
	\draw (3,4) circle[radius=1.5];
	\draw[dotted] (2.7,.1) -- (1.575,3.533);
	\draw[dotted] (2.7,.1) -- (4.338,3.3203);
	\node at (2.2,3) {$H$};
	\draw[->,rotate around={20:(2.6,4.2)}] (2.6,4.2,0)--(2.6,4.2,1);
	\draw[->,rotate around={20:(2.6,4.2)}] (2.6,4.2,0)--(3.4,4.2,0);
	\draw[->,rotate around={20:(2.6,4.2)}] (2.6,4.2,0)--(2.6,5,0);

	\filldraw (2.6,4.2) circle[radius=1pt] node[below right] {$\psi(p)$};
	\draw[->,dashed] (-.8,4.2) .. controls (0,4.8) and (1,4.6) .. node[above right] {$Q(X)$} (2.6,4.2);
	\draw[->,dashed] (2.6,4.2) -- node[right] {$q$} (2.7,0.1);
\end{tikzpicture}

\caption{Configuration of a Cosserat continuum using principal fibre bundles.
$p=(X,S)\in\mathcal{B}\times H=\mathcal{P}$ denotes an arbitrary
element of the bundle, which gets mapped to $\psi(p)=(\kappa(X),Q(X)S)\in\psi(\mathcal{P})\subset G$.}

\end{figure}
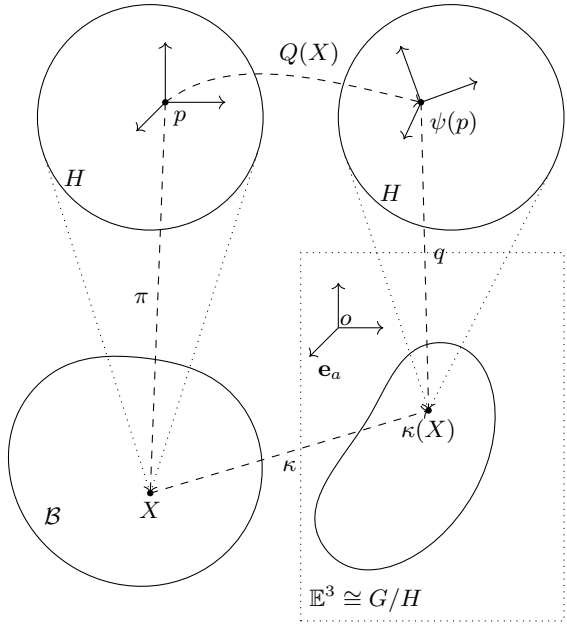

In this article we follow a slightly different approach to model Cosserat
continua by enlarging the material manifold $\mathcal{B}$ to $\mathcal{P}=\mathcal{B}\times H$,
where $H=\mathrm{O(3)}$ is the proper orthogonal group \citep{Epstein1998}.
This way $\mathcal{P}$ becomes a (trivial) principal fibre bundle
with structure group $H$ over the base space $\mathcal{B}$ (also
called the macromedium in this setting). It is instructive to think
of elements $p=(X,h)\in\mathcal{P}=\mathcal{B}\times H$ as infinitesimal
rigid bodies, where $X$ labels their centre of mass, and $h$ describes
their orientation and chirality, e.g. via a body frame. The right
action of $H$ on $\mathcal{P}$, given by $p=(X,h)\mapsto p\cdot k=(X,hk)$,
can be thought of as a change of body frame and a change of chirality
if $\det k=-1$. The projection map $\pi:\mathcal{P}\to\mathcal{B},(X,h)\mapsto X$
identifies the centre-of-mass label of a configuration of a material
point, while a section $s:\mathcal{B}\to\mathcal{P}$ can be thought
of as a specification of the orientation of the material particles.
Since a configuration of the base is given by a map from $\mathcal{B}$
to $\mathbb{E}^{3}$, a configuration of $\mathcal{P}$ should be
a map $\psi$ from $\mathcal{P}$ to an $H$-bundle over $\mathbb{E}^{3}$.
This bundle is usually taken to be the bundle of orthonormal (with
respect to the standard inner product on $\mathcal{E}$) frames $\mathcal{F}$
over $\mathbb{E}^{3}$. Nevertheless, it is convenient to single out
an origin $o\in\mathbb{E}^{3}$ and a Cartesian reference frame $\mathbf{e}_{i}\in\mathcal{E}$,
this way $\mathcal{F}$ can be identified with the group of isometries
(generated by translations, rotations and reflections) $G=\mathrm{E(3)}$
of $\mathbb{E}^{3}$. The identity element of $G$ will correspond
to the reference frame which also defines the positive orientation.
The bundle structure on $G$ is given by the quotient map $q:G\to G/H$.
The subgroup of orientation-preserving isometries of $\mathbb{E}^{3}$
will be denoted by $G^{+}=\mathrm{SE(3)}$.

\begin{figure}
\begin{tikzcd}[column sep=huge, row sep=large]	{\mathcal{P}} & G \\ 	{\mathcal B} & {G/H} 	\arrow["{\psi }", from=1-1, to=1-2] 	\arrow["{\pi }"', from=1-1, to=2-1] 	\arrow["q", from=1-2, to=2-2] 	\arrow["{\kappa }"', from=2-1, to=2-2] \end{tikzcd}

\caption{Commutative diagram corresponding to the bundle map $\psi:\mathcal{P}\to G$.}

\end{figure}

The configuration map $\psi:\mathcal{P}\to G$ should induce a configuration
$\kappa:\mathcal{B}\to G/H$ of the macromedium and respect the rigidity
of the microstructure. The first condition can be stated simply as:

\begin{equation}
q(\psi(p))=\kappa(\pi(p))\quad\forall p\in\mathcal{P}\label{eq:ind_conf}
\end{equation}
 The second condition means that for any $p\in\mathcal{P},h\in H$,
a rotated (and perhaps reflected) body frame $p\cdot h$ should get
mapped to the rotated (and perhaps reflected) frame $\psi(p)\cdot h$.
Therefore:
\begin{equation}
\psi(p\cdot h)=\psi(p)\cdot h\quad\forall p\in\mathcal{P},h\in H\label{eq:bund_map_equiv}
\end{equation}
Equation (\ref{eq:bund_map_equiv}) expresses that $\psi$ is an $H$-equivariant
map: in a certain sense, it is the mathematical manifestation of the
rigidity of the material particles. Properties (\ref{eq:ind_conf})
and (\ref{eq:bund_map_equiv}) imply that $\psi$ is an $H$-bundle
map between $\mathcal{P}$ and $G$ \citep{Epstein2007,Binz1998}.
Let us again choose a reference configuration $\psi_{0}:\mathcal{P}\to G$,
then by identifying $\mathcal{P}$ with $\psi_{0}(\mathcal{P})\subset G$
so that $\psi_{0}\equiv\mathrm{Id}$, we assume from now on without
loss of generality that $\mathcal{P}$ is a subbundle of $G$. In
what follows, we will often use the following representation of elements
of $G$ by $4\times4$-matrices:

\begin{equation}
p=\begin{bmatrix}1 & 0\\
x & S
\end{bmatrix}\label{eq:eucl_rep}
\end{equation}
where $x\in G/H\cong\mathbb{R}^{3}$ and $S\in\mathrm{O(3)}$. The
group element $p$ in (\ref{eq:eucl_rep}) corresponds to the orthonormal
frame based at the point $o+x_{a}\mathbf{e}_{a}$ with basis vectors
$S_{ab}\mathbf{e}_{a}$. Now let $\psi:\mathcal{P}\to G$ be an $H$-bundle
map, then by the above stated properties it must be of the form

\begin{equation}
\psi(p)=\begin{bmatrix}1 & 0\\
y(x) & Q(x)S
\end{bmatrix}\label{eq:bundle_map}
\end{equation}
for some functions $y:\mathcal{B}\to\mathbb{R}^{3}$ and $Q:\mathbb{\mathcal{B}}\to H$
representing the deformation of the macromedium and the microstructure
respectively, recovering the usual setting oulined in the second paragraph
of this section. Since we are concerned with deformations that are
continuously attainable from the identity, we will restrict $Q(x)$
to have determinant $+1$ so that $Q:\mathcal{B}\to\mathrm{SO}(3)$
for the remainder of this article.

We will denote the Lie algebra of $G$ (which is the same as the Lie
algebra of $G^{+}$) by $\mathfrak{g}=\mathfrak{se}(3)$, and the
Lie algebra of $H$ by $\mathfrak{h}=\mathfrak{so}(3)$. The matrix
representation (\ref{eq:eucl_rep}) induces a matrix representation
of $\mathfrak{g}$ given by:

\begin{equation}
w=\begin{bmatrix}0 & 0\\
u & \Phi
\end{bmatrix}\label{eq:lie_alg_rep}
\end{equation}
where $u\in\mathbb{R}^{3}$ corresponds to an infinitesimal translation
while $\Phi\in\mathfrak{h}$ is a $3\times3$ antisymmetric matrix
corresponding to an infinitesimal rotation, which is usually identified
with an axial vector $\varphi\in\mathbb{R}^{3}$. The Lie bracket
on $\mathfrak{g}$ is given by the usual matrix commutator: for $w,z\in\mathfrak{g}$,
we have $[w,z]=wz-zw$, and the adjoint action of $G$ on $\mathfrak{g}$
is given by matrix conjugation: for $g\in G,w\in\mathfrak{g}$, $\mathrm{Ad}_{g}w=gwg^{-1}$.
Note also that the Lie algebra splits as $\mathfrak{g}=\mathfrak{m}\oplus\mathfrak{h}$,
where $\mathfrak{m}$ is the $\mathrm{Ad}H$-invariant subspace of
infinitesimal translations (which can be identified with every tangent
space of $\mathbb{E}^{3}$), meaning that $\mathbb{E}^{3}$ as a homogeneous
space $G/H$ is reductive \citep{Kobayashi1996}. This slightly technical
property of $G$ and $H$ will allow us to separate deformation and
incompatibility measures into translational $\mathfrak{m}$ and rotational
$\mathfrak{h}$ parts; otherwise, we could only treat them together
as $\mathfrak{g}$-valued objects \citep{Sharpe1997,Wise2010} (see
below).

\section{Theory of Strain\label{sec:Theory-of-Strain}}

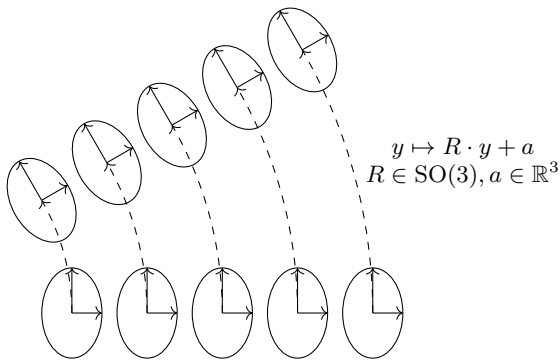
\begin{figure}
\begin{tikzpicture}
	\foreach \x in {3,...,7} {\draw (\x,0) ellipse [x radius=0.4, y radius=0.6]; \draw[->] (\x,0) -- (\x+0.4,0); \draw[->] (\x,0) -- (\x,.6); \draw[->,dashed] (\x,0) arc [start angle=0, end angle=30,radius=\x];}
	\begin{scope}[rotate=30]
		\foreach \x in {3,...,7} {\draw (\x,0) ellipse [x radius=0.4, y radius=0.6]; \draw[->] (\x,0) -- (\x+.4,0); \draw[->] (\x,0) -- (\x,0.6);}
	\end{scope}
	\node[align=center] at (8.2,2) {$y\mapsto R\cdot y+a$\\$R\in\mathrm{SO}(3),a\in\mathbb{R}^3$};
\end{tikzpicture}

\caption{Rigid transformation of a Cosserat continuum. Observe that the material
particles rotate together with spatial points, therefore spatial and
microstructural rotations are not independent.}

\end{figure}

In continuum mechanics, one is interested in configurations up to
rigid transformations. The building block of any continuum theory
is a strain measure which captures the deviation of a mapping of a
continuum body from a rigid transformation. It should be nonzero if
and only if $\psi$ differs from a global rigid body transformation.
In the current framework, rigid motions are implemented by left multiplication
$L_{g}:\mathcal{P}\to G,p\mapsto g\cdot p$ by a constant group element
$g\in G^{+}$. Note that such a transformation acts simultaneously
on the base (macromedium) and fibres (micromedium), therefore under
a superimposed rigid transformation $y\mapsto R\cdot y+a$ for $R\in\mathrm{SO(3)},a\in\mathbb{R}^{3}$
the microstructure directors rotate together with the basepoint by
$R$ \citep{Maitra2019}. This assumption, one of the most important
in Cosserat theory, is often stated in the literature as the objectivity
of microstructure directors \citep{Le1998}.

In view of the above considerations, we consider the quantity:
\begin{equation}
E=\psi^{-1}d\psi-p^{-1}dp=\psi^{*}\omega-\omega\label{eq:finite_strain}
\end{equation}
where $\omega$ is the Maurer-Cartan form on the group $G$ and $\psi^{*}\omega$
denotes the pullback of $\omega$ along the map $\psi$. The Maurer-Cartan
form satisfies the well-known Maurer-Cartan structure equations:

\begin{equation}
d\omega+\omega\wedge\omega=0\label{eq:mc_str_eq}
\end{equation}
If we write $\psi(p)=k(p)\cdot p$ for a function $k:\mathcal{P}\to G$,
then substituting into equation (\ref{eq:finite_strain}) yields:

\begin{equation}
E=p^{-1}\left(k^{-1}dk\right)p+p^{-1}dp-p^{-1}dp=p^{-1}\left(k^{-1}dk\right)p\label{eq:finite_strain_explain}
\end{equation}

\begin{figure}
\begin{tikzpicture}[scale=1.5]
	\foreach \x in {3,...,7} {\draw (\x,0) ellipse [x radius=0.4, y radius=0.6]; \draw[->] (\x,0) -- (\x+0.4,0); \draw[->] (\x,0) -- (\x,.6);}
	\foreach \x in {3,...,7} {
				\filldraw (\x,2) circle [radius=1pt];
				\draw[rotate around={30:(\x,2)}] (\x,2) ellipse [x radius=0.4, y radius=0.6];
				\draw[->,rotate around={30:(\x,2)}] (\x,2) -- (\x+.4,2);
				\draw[->,rotate around={30:(\x,2)}] (\x,2) -- (\x,2.6);
	}
	\foreach \x in {3,...,6} {\draw[dotted] (\x,2)--(\x+1,2); \draw (\x+.25,2) arc [start angle=0,end angle=30, radius=.25];}
\end{tikzpicture}\caption{Unlike magnetic systems or the $\mathrm{O}(n)$ model in statistical
field theory \citep{Chaikin1995}, rotation of material particles
of a Cosserat continuum independent of spatial rotations induces a
deformation \citep{Maitra2019}. This can be inferred by looking at
the change in the angle between microstructure directors and vectors
separating the centres of mass of material particles.}
\end{figure}
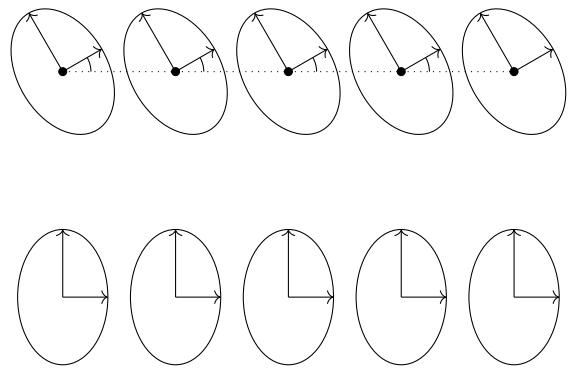
Equation (\ref{eq:finite_strain_explain}) shows that $E$ is identically
zero if and only if $k:\mathcal{P}\to G$ is a constant function,
which is equivalent to $\psi$ being a global rigid transformation.
The deformation measure introduced in (\ref{eq:finite_strain}) is
analogous to the classical Green-Lagrangian strain which is defined
as the difference between the pullback of the spatial metric and the
reference metric \citep{Marsden1994}. In this case, the role of the
metric as the indicator of deformation is replaced by the Maurer-Cartan
form, which in turn can be viewed as a certain kind of connection
(a Cartan connection) on the bundle $\mathcal{P}$ (for an in-depth
presentation of Cartan connections, see \citep{Sharpe1997}). In the
appendix, we provide an informal motivation as to why the strain measure
is a difference of connections. While the appearance of the Maurer-Cartan
form in (\ref{eq:finite_strain}) might initially seem a bit surprising
and perhaps difficult to grasp, it has been used implicitly and unconsciously
in numerous contexts and applications related to elasticity and soft
matter. For instance, strain measures in beam or plate theories \citep{Reissner1981,Simo1988,Simo1985,Ellis2010}
(also introduced based on left-invariance under rigid transformations)
are in fact closely related to (\ref{eq:finite_strain}).

Let us now compute $E$ explicitly. Using (\ref{eq:eucl_rep}) and
(\ref{eq:bundle_map}) we have:

\begin{alignat}{1}
\omega=p^{-1}dp= & \begin{bmatrix}1 & 0\\
-S^{T}x & S^{T}
\end{bmatrix}\begin{bmatrix}0 & 0\\
dx & dS
\end{bmatrix}\label{eq:mc_coord}\\
= & \begin{bmatrix}0 & 0\\
S^{T}dx & S^{T}dS
\end{bmatrix}\nonumber 
\end{alignat}

Similarly:

\begin{gather}
\psi^{-1}d\psi=\begin{bmatrix}1 & 0\\
-S^{T}Q^{T}y & S^{T}Q^{T}
\end{bmatrix}\begin{bmatrix}0 & 0\\
dy & dQ\cdot S+QdS
\end{bmatrix}=\label{eq:pullback_mc}\\
=\begin{bmatrix}0 & 0\\
S^{T}Q^{T}dy & S^{T}\left(Q^{T}dQ\right)S+S^{T}dS
\end{bmatrix}\nonumber 
\end{gather}

Substracting (\ref{eq:mc_coord}) and (\ref{eq:pullback_mc}) we find:

\begin{equation}
E=\begin{bmatrix}0 & 0\\
S^{T}\left(Q^{T}dy-dx\right) & S^{T}\left(Q^{T}dQ\right)S
\end{bmatrix}\label{eq:finite_strain_coords}
\end{equation}

Thus we recover the usual finite translational $\mathbf{Q}^{T}dy-dx$
and rotational $\mathbf{Q}^{T}d\mathbf{Q}$ strain measures of the
Cosserat solid, where $\mathbf{Q}(x)$ is the usual notation for the
proper orthogonal rotation tensor describing the deformation of microstructure
directors \citep{Pietraszkiewicz2009,Bohmer2015,Maugin1998}. The
factors of $S$ illustrate that $E$ is tensorial: its components
change in representations of $\mathrm{O(3)}$ under a change of section
$S$, with the rotational strain measure unchanged if $S=-I$ is an
inversion, indicating that it is pseudovector valued. As we will see
shortly, $E$ is in fact a vector valued $1$-form over $\mathcal{B}$.

To obtain a consistent linearization of $E$, we consider infinitesimal
deformations of the continuum, which are vector fields $V$ on $\mathcal{P}$
such that their flows preserve the bundle structure \citep{GayBalmaz2009}
(i.e. they are $H$-bundle maps). This puts the following constraint
on the vector field: 
\begin{equation}
V(ph)=V(p)\cdot h\quad\forall p\in\mathcal{P},h\in H\label{eq:inf_def_vec}
\end{equation}
As $V$ is a vector field on a Lie group $G$, we can represent it
by a Lie algebra valued function $\xi:\mathcal{P}\to\mathfrak{g}$
such that $V(p)=p\cdot\xi(p)$, or equivalently $\xi(p)=\iota_{V}\omega|_{p}$,
where $\iota$ denotes the interior product. Note also that $\xi(p)$
satisfies:

\begin{equation}
\xi(ph)=h^{-1}\xi(p)h=Ad\left(h^{-1}\right)\xi(p)\label{eq:inf_def_lie_alg}
\end{equation}
Hence $\xi$ is a section of the associated vector bundle $W=\mathcal{P}\times_{H}\mathfrak{g}$
transforming in the adjoint representation of $H$ on $\mathfrak{g}$.
This vector bundle is what \citet{Schaefer1967} calls the space of
motors. The Lie derivative, by definition, is the infinitesimal version
or first order approximation of the difference $\psi^{*}-\mathrm{Id}$
along the flow map of a vector field. Therefore, the linearization
$e$ of $E$ is obtained by taking the Lie derivative of $\omega$
along the vector field $V$, which measures the infinitesimal change
of $\omega$ along the vector field $V$ \citep{Marsden1994,Hehl2007}.
This is further elaborated in the appendix. Using Cartan's magic formula
\citep{Frankel2011} and the Maurer-Cartan structure equations (\ref{eq:mc_str_eq})
we get:

\begin{alignat}{1}
e & \triangleq\mathcal{L}_{V}\omega=d(\iota_{V}\omega)+\iota_{V}d\omega\nonumber \\
 & =d\xi-\iota_{V}(\omega\wedge\omega)\nonumber \\
 & =d\xi-\xi\wedge\omega+\omega\wedge\xi\label{eq:inf_strain_lie}\\
 & =d\xi+\mathrm{ad}_{\omega}\xi\nonumber 
\end{alignat}

If $V$ is not an infinitesimal displacement but a velocity vector
field corresponding to a motion of the Cosserat continuum, then (\ref{eq:inf_strain_lie})
defines the strain rate. One may also argue that (\ref{eq:inf_strain_lie})
is in fact a more fundamental strain measure than (\ref{eq:finite_strain_coords})
because it does not assume the existence of an arbitrary reference
configuration. In addition, it is closer in spirit to differential
geometry, where every meaningful operation or comparison can only
be done locally, and global results can be obtained by integration.
In the next section, an explicit demonstration of this viewpoint will
be shown by taking $e$ to be a general Lie algebra valued $1$-form
not necessarily coming from a vector field $V$, leading to the appearance
of topological defects. Moreover, every reasonable elastic deformation
can be decomposed into a sequence of infinitesimal deformations, consequently
a finite strain measure may be obtained by integrating (\ref{eq:inf_strain_lie})
along the motion \citep{Post1956,Romano2014}. In the autonomous case
when the vector field is independent of time parametrizing the motion,
this integration gives the exponential of the Lie derivative operator,
which results in the pullback operation along the flow of the vector
field $V$, consistent with (\ref{eq:finite_strain_coords}).

Let us now compute $e$ in Cartesian coordinates on $\mathcal{B}$.
Formally, this is obtained by pulling back $e$ to $\mathcal{B}$
via the section corresponding to $S=I$. Locally $\xi$ is represented
by a $\mathfrak{g}$-valued function on $\mathcal{B}$, which we can
specify by a pair of infinitesimal displacement and microrotation
fields $u:\mathcal{B}\to\mathbb{R}^{3}$ and $\Phi:\mathcal{B}\to\mathfrak{so}(3)$
\footnote{This is one instance where we make use of the fact that $G/H$ is
a reductive space: in different coordinates, $\xi(x)$ transforms
to $h(x)^{-1}\xi(x)h(x)=\mathrm{Ad}_{h(x)^{-1}}\xi(x)$, so the translational
piece transforms correctly if $\mathrm{Ad}_{h(x)^{-1}}u(x)$ is also
in $\mathfrak{m}$. }. In Cartesian coordinates, since $\omega$ pulls back to the $\mathbb{R}^{3}$-valued
$1$-form $d\boldsymbol{x}$ (corresponding to the identity tensor
on $\mathcal{B}$) we get:

\begin{gather}
e=\begin{bmatrix}0 & 0\\
du & d\Phi
\end{bmatrix}-\begin{bmatrix}0 & 0\\
u & \Phi
\end{bmatrix}\wedge\begin{bmatrix}0 & 0\\
d\boldsymbol{x} & 0
\end{bmatrix}+\begin{bmatrix}0 & 0\\
d\boldsymbol{x} & 0
\end{bmatrix}\wedge\begin{bmatrix}0 & 0\\
u & \Phi
\end{bmatrix}=\nonumber \\
=\begin{bmatrix}0 & 0\\
du-\Phi d\boldsymbol{x} & d\Phi
\end{bmatrix}=\begin{bmatrix}0 & 0\\
\varepsilon & \tau
\end{bmatrix}\label{eq:inf_comp}
\end{gather}
Under the usual identification of elements $\mathfrak{so}(3)$ with
$\mathbb{R}^{3}$, $\Phi$ is identified with the axial vector field
$\boldsymbol{\varphi}:\mathcal{B}\to\mathbb{R}^{3}$ and the matrix
product $\Phi d\boldsymbol{x}$ becomes a vector cross product $\boldsymbol{\varphi}\times d\boldsymbol{x}$.
This way we recover the well-known infinitesimal strain measures of
the Cosserat continuum $d\boldsymbol{u}-\boldsymbol{\varphi}\times d\boldsymbol{x}$
and $d\boldsymbol{\varphi}$ \citep{Cosserat1909}\citep{Eringen1999}.
In components ($\epsilon_{ijk}$ denotes the Levi-Civita permutation
symbol, coming from the commutation relations of $\mathfrak{g}$):

\begin{alignat}{1}
\left(d\boldsymbol{u}-\boldsymbol{\varphi}\times d\boldsymbol{x}\right)_{j} & :=\varepsilon_{ij}dx^{i}=\left(\partial_{i}u_{j}-\epsilon_{ijk}\varphi_{k}\right)dx^{i}\label{eq:inf_coord_tr}\\
\left(d\boldsymbol{\varphi}\right)_{j} & :=\tau_{ij}dx^{i}=\left(\partial_{i}\varphi_{j}\right)dx^{i}\label{eq:inf_coord_rot}
\end{alignat}

Both $E$ and $e$ are one-forms with values in the bundle $W$: this
is due to the fact that both of them are essentially differences of
connections on $\mathcal{P}$, and as such are tensor-valued (in this
case $W$-valued) one-forms on $\mathcal{B}$. Therefore equation
(\ref{eq:inf_strain_lie}) can also be interpreted as the covariant
derivative $D\xi$ of $\xi$ with respect to the connection on $W$
induced by $\omega$, the connection coefficients in Cartesian coordinates
can be read off from (\ref{eq:inf_coord_tr}) and (\ref{eq:inf_coord_rot}).
This result was first obtained by \citet{Schaefer1967}. Hence the
infinitesimal strain measure can be equally viewed as a covariant
derivative of a section of $W$ or the Lie derivative of the Maurer-Cartan
form along a vector field on $G$. This also has an analogue in classical
elasticity, where infinitesimal strain is the Lie derivative $\mathcal{L}_{U}g$
of the metric $g$ along a vector field $U$, but can also be expressed
as $\frac{1}{2}\left(\nabla U+\left(\nabla U\right)^{T}\right)$ with
the aid of the Levi-Civita connection $\nabla$ corresponding to $g$
\citep{Romano2013}. It is also interesting to remark that the infinitesimal
strain measures (\ref{eq:inf_coord_tr})--(\ref{eq:inf_coord_rot})
have been recently derived in the context of symmetry breaking and
high energy physics \citep{Hirono2022} via a coset construction starting
from the Maurer-Cartan form on the Galilei group. Furthermore, Cartan
connections play an important role in the extended theories of general
relativity \citep{Hehl1995,Wise2010} (which have, in turn, also been
influenced by the theory of continua with microstructure).

It should be apparent from the development above that the Maurer-Cartan
form is the basic building block of almost any classical continuum
theory respecting locality and invariance under rigid body transformations
because it furnishes a set of differential invariants that can subsequently
be used to construct Lagrangians or free energies for conservative
systems in equilibrium. The most well-known example of this procedure
is the Frenet-Serret framing \citep{CarmoManfredoPerdigaodo2016}:
for a curve $\gamma:(a,b)\to\mathbb{E}^{3}\cong G/H$ embedded in
Euclidean space, the set of tangent, normal and binormal vectors give
an adapted frame that provides a lift $\tilde{\gamma}:(a,b)\to G$
of the curve from the Euclidean (homogeneous) space to the Euclidean
group . The curvature and torsion (not to be confused with curvature
and torsion of a connection \citep{Delphenich2016} in later sections)
of the curve are nothing else but the components of the pullback of
the Maurer-Cartan form on $G$ via the lift $\tilde{\gamma}$. The
simplest resulting elasticity theory of curves in two dimensions is
the famous theory of the elastica dating back to Euler \citep{Levien:EECS-2008-103}.
The classification of submanifolds of general homogeneous spaces has
led to the deep and beautiful method of moving frames \citep{Clelland2017,Olver1995},
pioneered by \'Elie Cartan, where the Maurer-Cartan form plays a fundamental
role.

\section{Compatibility Conditions\label{sec:Compatibility-Conditions}}

If $e$ is an $W$-valued one-form, a natural question to ask is that
when it is compatible with a displacement field, i.e. whether there
exist $u,\Phi$ satisfying (\ref{eq:inf_comp}). Necessary conditions
can be elegantly obtained by means of exterior calculus, utilizing
the exterior covariant derivative induced on $W$-valued differential
forms by the connection $\omega$ \citep{Schaefer1967}. This is done
as follows: any $W$-valued $p$-form $\alpha$ can be equivalently
characterized as a $\mathfrak{g}$-valued $p$-form on $\mathcal{P}$
that is $H$-equivariant and horizontal \footnote{A $\mathfrak{g}$-valued differential $k$-form $\alpha$ on $\mathcal{P}$
is $H$-equivariant if $R_{h}^{*}\alpha=\mathrm{Ad}_{h^{-1}}\alpha$
and horizontal if $\alpha(v_{1},\dots,v_{k})=0$ whenever one of the
$v_{i}$-s is belong to the kernel of the differential of the projection
map $\pi:\mathcal{P}\to\mathcal{B}$. These conditions ensure that
$\alpha$ descends to a $W$-valued $p$-form on $\mathcal{B}$}. Now the exterior covariant derivative $D\alpha$ is the $\mathfrak{g}$-valued
$p+1$-form on $\mathcal{P}$ defined as:

\begin{equation}
D\alpha:=d\alpha+\mathrm{ad}_{\omega}\alpha=d\alpha+\omega\wedge\alpha-(-1)^{p}\alpha\wedge\omega\label{eq:ext_cov_deriv_def}
\end{equation}
It can be verified that (\ref{eq:ext_cov_deriv_def}) is again an
equivariant and horizontal form, hence descends to a $W$-valued $p+1$-form
on $\mathcal{B}$. The compatibility conditions follow from the fact
that the connection is flat. One can show that:

\begin{equation}
DD\alpha=(d\omega+\omega\wedge\omega)\wedge\alpha-\alpha\wedge(d\omega+\omega\wedge\omega)=0\label{eq:ext_cov_deriv_flat}
\end{equation}
Therefore a necessary \footnote{Whether or not it is sufficient depends on the topology of the body
manifold.} condition for a $W$-valued $1$-form $e$ to be integrable (i.e
to be of the form $D\xi$ for some section $\xi$ of $W$) is 
\begin{equation}
De=de+\omega\wedge e+e\wedge\omega=0\label{eq:str_integr}
\end{equation}
In coordinates (with a slight abuse of notation as $\tau$ here corresponds
to an $\mathfrak{so}(3)$-valued $1$-form): 

\begin{gather}
De=\begin{bmatrix}0 & 0\\
d\varepsilon & d\tau
\end{bmatrix}+\begin{bmatrix}0 & 0\\
d\boldsymbol{x} & 0
\end{bmatrix}\wedge\begin{bmatrix}0 & 0\\
\varepsilon & \tau
\end{bmatrix}+\begin{bmatrix}0 & 0\\
\varepsilon & \tau
\end{bmatrix}\wedge\begin{bmatrix}0 & 0\\
d\boldsymbol{x} & 0
\end{bmatrix}\nonumber \\
=\begin{bmatrix}0 & 0\\
d\varepsilon+\tau\wedge d\mathbf{x} & d\tau
\end{bmatrix}=0\label{eq:comp_compute}
\end{gather}
Using the definition of the exterior derivative and the correspondence
between the action of $\mathfrak{so}(3)$ on vectors and the vector
cross product we obtain \citep{Eringen1999}:

\begin{alignat}{1}
\left(\partial_{k}\varepsilon_{ij}-\partial_{i}\varepsilon_{kj}\right)-\left(\epsilon_{kjl}\tau_{il}-\epsilon_{ijl}\tau_{kl}\right) & =0\label{eq:comp_tr}\\
\partial_{i}\tau_{jk}-\partial_{j}\tau_{ik} & =0\label{eq:comp_rot}
\end{alignat}
Any nonzero quantities on the right-hand side of (\ref{eq:comp_tr})
and (\ref{eq:comp_rot}) can be interpreted as measures of incompatibility.
On the right hand side of (\ref{eq:comp_tr}) there can be a displacement-valued
two-form $T$, usually associated with torsion, while on the right-hand
side of (\ref{eq:comp_rot}) a microrotation-valued two-form $\Omega$,
associated with curvature \citep{Lazar2010}. Therefore, in components,
in general we have:

\begin{alignat}{1}
\left(\partial_{k}\varepsilon_{ij}-\partial_{i}\varepsilon_{kj}\right)-\left(\epsilon_{kjl}\tau_{il}-\epsilon_{ijl}\tau_{kl}\right) & =T_{ijk}\label{eq:comp_tr-1}\\
\partial_{i}\tau_{jk}-\partial_{j}\tau_{ik} & =\Omega_{ijk}\label{eq:comp_rot-1}
\end{alignat}
They can be viewed together as a more general $W$-valued $2$-form,
$J$, satisfying the Bianchi identity (\ref{eq:bianchi}) \citep{Lazar2007}.

\begin{alignat}{1}
De & =J\label{eq:curv_def}\\
DJ & =0\label{eq:bianchi}
\end{alignat}
For the finite strain measure $E$, a similar compatibility condition
can be derived based on the Maurer-Cartan structure equations: since
\[
E+\omega=\psi^{*}\omega,
\]
the integrability condition reads:

\begin{equation}
d(E+\omega)+(E+\omega)\wedge(E+\omega)=dE+E\wedge\omega+\omega\wedge E=0\label{eq:finite_comp}
\end{equation}

In general, one can model a Cosserat solid with defects by an abstract
principal $H$-bundle $\mathcal{P}$ equipped with a Cartan connection
$\eta$, that is, a $\mathfrak{g}$-valued $1$-form on $\mathcal{P}$
satisfying the following properties \citep{Sharpe1997}:
\begin{enumerate}
\item $R_{h}^{*}\eta=Ad\left(h^{-1}\right)\eta\quad\forall h\in H$, i.e.
$\eta$ is $H$-equivariant.
\item $\eta\left(p\cdot\varsigma\right)=\varsigma\quad\forall\varsigma\in\mathfrak{h}$,
where $p\cdot\varsigma=\frac{d}{dt}\bigg|_{t=0}\left(p\cdot\exp(t\varsigma)\right)$.
\item $\eta|_{p}:T_{p}\mathcal{P}\to\mathfrak{g}$ is a linear isomorphism
for all $p\in\mathcal{P}$.
\end{enumerate}
Albeit rather formally (for an intuitive interpretation of the above
properties of $\eta$, see e.g. \citep{Wise2010}), these three conditions
capture and generalize the main properties of the Maurer-Cartan form
on the principal $H$-bundle $G\to G/H$. The crucial difference is
that the Maurer-Cartan structure equations do not necessarily hold:
the two-form $\Theta=d\eta+\eta\wedge\eta$ is in general nonzero.
This quantity is called the curvature of the Cartan connection, representing
the incompatibility of the underlying Cosserat solid. Note that this
curvature is more general than the curvature of a linear connection
because it takes values in the Lie algebra of the Euclidean group,
not a subgroup of the general linear group. In some sense, one can
view $\Theta$ as the ``unification'' of torsion and curvature into
a single object, whose significance is highlighted by the fact that
rotational and translational defects are intimately coupled in Cosserat
solid as expressed by (\ref{eq:comp_tr-1}) and (\ref{eq:comp_rot-1}).

Modelling bodies with continuous distributions of topological defects
(dislocations, disclinations, etc.) as abstract manifolds equipped
with an extra structure incompatible with Euclidean space has a long
and distinguished history, starting from the work of \citet{Nye1953},
\citet{Bilby1955}, \citet{Kroner1959}, \citet{Kondo1964} and others
in the 1950s and 1960s. These authors have independently discovered
that the Burgers vector density of continuous distributions of dislocations
can be mapped to the torsion of a certain affine connection on the
material manifold. It has also been realized that continuous distributions
of disclinations and point defects are associated with the curvature
(in the usual Riemannian sense) and the non-metricity of an affine
connection on the material manifold, see e.g. \citep{Yavari2012,Yavari2013}
for a modern exposition.

While the compatibility conditions and stress fields around defects
for Cosserat solids have been obtained by various authors, the geometric
origin of incompatibilities remained unclear. Based on the above discussion
and inspired by \citep{Yavari2012}, we propose a simple model of
an incompatible Cosserat solid as a principal $H$-bundle $\mathcal{P}$
equipped with a non-flat Cartan connection $\eta$. The finite strain
measure $E$ becomes $\psi^{*}\omega-\eta$ and (\ref{eq:finite_comp})
changes to:

\begin{equation}
d(E+\eta)+(E+\eta)\wedge(E+\eta)=D_{\eta}E+\Theta=0\label{eq:finite_comp_2}
\end{equation}
where $D_{\eta}$ is the exterior covariant derivative with respect
to the connection $\eta$. This definition of strain as a difference
of connections also sheds light on the appearance of the contorsion
tensor in previous works on topological defects in general relativity
and complex media \citep{Randono2011,Bohmer2021}. In addition, with
the aid of the theory of stress and and constitutive relations given
in the following sections, one can calculate the stress fields around
defects in Cosserat solids \citep{Kessel1970}.

\section{Theory of Stress and Balance Laws\label{sec:Theory-of-Stress}}

Most treatments of classical elasticity derive the governing equations
of elasticity via the following train of thought \citep{Landau1986}:
interactions between material particles are assumed to be short-ranged,
which, together with the action-reaction principle, implies the existence
of a second-rank stress tensor through Cauchy's tetrahedron argument.
Equations of motion are then obtained by postulating the balance of
linear and angular momentum on each subbody of a macroscopic body,
with the former yielding Cauchy's momentum equations while the latter
the symmetry of the stress tensor upon localization. However, there
are a number of difficulties encountered when trying to extend this
method to complex materials \citep{Maugin2010}.

First, the above assumptions restrict the nature of boundary interactions
between subbodies to be only force-like depending linearly on the
normal of the boundary surface. However, one can envisage cases when
``higher-order'' forces corresponding to media sensitive to higher
gradients in displacements or the curvature of boundary surfaces \citep{Maugin2010,dellIsola2020}.
In the case of complex materials, torques or even torque dipoles could
be transmitted through boundaries. In general, it is unclear what
kind of balance laws one should postulate for these quantities.

Second, the relationship between the balance of angular momentum and
the symmetry of stress tensor is quite mysterious: in Cauchy elasticity
\citep{Landau1986}, this balance law is only used to argue that the
stress tensor is symmetric, but subsequently neglected in the solution
of any practical problem \citep{Truesdell1968}. Furthermore, if material
points in a Cauchy continuum are assumed to be structureless point
masses only possessing translational degrees of freedom, why does
the balance of angular momentum constitute an independent equation?
It is in stark contrast with the Newtonian mechanics of point masses,
where Newton's second law is the only equation needed to work out
the motion of a point mass.

Finally, from a differential geometric standpoint, the vector-valued
integrals in balance laws are ill-defined because in non-flat spaces
one cannot identify distant tangent spaces unambiguously.

To address these problems, we construct the theory of stress and balance
laws from another perspective, based on the principle of virtual work
\citep{Eugster2017}. Even though this approach is much less popular
and appreciated than Cauchy's, it also has a long and distinguished
history, essentially originating from Piola who extended the ideas
of d'Alembert to the mechanics of continua \citep{dellIsola2020}.
The fundamental assumption is that applying a rigid transformation
to a body requires no work, which motivates the introduction of strain
as a measure of deviation from a rigid transformation (in classical
elasticity, it is the change in the metric along the deformation,
while in the present case it is the change in the connection as argued
in Section III). Stress is defined as a dual quantity to strain: the
duality pairing gives the virtual work done by stress during an infinitesimal
variation of configuration. This variation or virtual displacement
is represented by a vector field on the current configuration respecting
the imposed boundary conditions (in classical elasticity, it is a
vector field on the ambient space, while for Cosserat solids it is
a vector field exactly analogous to $V$ in (\ref{eq:inf_def_vec})).
Equations of motion are obtained by a version of the d'Alembert's
principle which states that the total virtual work is zero for any
admissible virtual displacement field. This method answers all the
above concerns as follows.

First, in the variational formalism, stress has exactly as many degrees
of freedom as the model of the continuum has: for rotational degrees
of freedom, one has moment stresses, and for higher-gradient forces
additional ``hyperstresses'' \citep{Germain1973,Germain2020,dellIsola2020}.
Balance laws are obtained from a single application of the principle
of virtual work.

Second, the duality between stress and strain clearly highlights the
number of degrees of freedom in a model and the underlying symmetries.
For example, in classical elasticity the infinitesimal strain $\frac{1}{2}\left(\partial_{i}u_{j}+\partial_{j}u_{i}\right)$
is an element of the vector space of \textsl{symmetric }second-rank
tensors, therefore the stress tensor, being an element of the dual
vector space, is also \textsl{automatically} symmetric. The symmetry
of the stress tensor has thus been identified as a consequence of
the number of degrees of freedom in a Cauchy continuum (because the
strain only depends on the displacement vector) and the axiom that
rigid transformations require no work \citep{Eugster2017}. This way
in Cauchy elasticity the only balance law following from the principle
of virtual work is Cauchy's momentum equation.

Finally, the principle of virtual work involve scalar-valued integrals
only which are well-defined on any smooth manifold.

Motivated by the above discussion and inspired by other geometric
treatments of elasticity \citep{Kanso2007,Rashad2023,Frankel2011},
for Cosserat solids we therefore define stress $\Sigma$ as an $\tilde{W}^{*}$-valued
$2$-form \footnote{To be precise, $\Sigma$ is a pseudoform as it depends on the orientation
of space, but this distinction is not cruical here, see \citep{Frankel2011}
for more details.} on the current configuration $\psi(\mathcal{P})$, where $\tilde{W}^{*}$
is the dual vector bundle of $\tilde{W}=\psi(\mathcal{P})\times_{H}\mathfrak{g}$
(this is the space of dual motors in the language of \citep{Schaefer1967}).
The intuitive reason behind this definition is that (Eulerian) velocities
are sections of $\tilde{W}$ (analogously as in (\ref{eq:inf_def_vec})),
and the natural duality pairing between stress and velocity gives
a scalar-valued $2$-form that can be integrated on the current configuration
to obtain the power of stresses exerted on surfaces inside the body.
(More generally, stress can be defined in $n$ dimensions as an $n-1$
form taking values in a vector bundle dual to velocities/virtual displacements,
and thus can be integrated on any $n-1$-submanifold to give the rate
of power on a hypersurface, see e.g. \citep{Gronwald1997,Frankel2011}).
While this would initially suggest that $\Sigma$ is described by
three indices (one bundle index and two form indices), one usually
reduces this to two indices overall to obtain the second rank Cauchy
stress tensor in the presence of a metric and a volume form by exploiting
the isomorphism provided by the Hodge duality between $2$-forms and
$1$-forms, i.e. by associating area elements with corresponding normal
vectors. It turns out that there is a more general correspondence
between $n-1$-forms and vectors afforded by the interior product
which only requires a volume form, so the existence of a metric structure
is not required.

The duality pairing between stress and infinitesimal strain or strain
rate gives a scalar-valued $3$-form on $\tilde{\mathcal{B}}=\varphi(\mathcal{B})$,
which can be integrated over any $3$-dimensional submanifold of $\tilde{\mathcal{B}}$
to give the virtual work or the rate of work done by stresses, respectively
\citep{Kanso2007}. This pairing can be defined for general $p$-
and $q$-forms taking values in $\tilde{W}^{*}$ and $\tilde{W}$,
outputting a scalar-valued $p+q$-form on $\mathcal{B}$. For forms
$\mu\otimes\alpha$ and $s\otimes\beta$, where $\mu$ and $s$ are
sections of $\tilde{W}^{*}$ and $\tilde{W}$, $\alpha$ and $\beta$
are $p$ and $q$-forms on $\mathcal{B}$, it is given by $\langle\mu\otimes\alpha,s\otimes\beta\rangle:=\langle\mu,s\rangle\alpha\wedge\beta$,
then extended linearly to all forms (here $\langle\mu,s\rangle$ is
the natural pairing of sections of $\tilde{W}$ and $\tilde{W}^{*}$).

Let us compute this pairing explicitly for the stress $2$-form $\Sigma$
and the infinitesimal strain $1$-form $e$. We choose a basis $\left\{ \mathbf{v}_{i},\mathbf{r}_{i}\right\} $
of $\mathfrak{g}$, where the $\left\{ \mathbf{v}_{i}\right\} $ generate
infinitesimal translations (spanning the subspace $\mathfrak{m}$)
and the $\left\{ \mathbf{r}_{i}\right\} $ generate infinitesimal
rotations (spanning the subspace $\mathfrak{h}$). Suppose that in
local Cartesian coordinates $x_{i}$, $i=1,2,3$, $e$ is represented
by the pair of $\mathfrak{m}$- and $\mathfrak{h}$-valued $1$-forms
$\varepsilon_{ij}\mathbf{v}_{j}dx_{i}$ and $\tau_{ij}\mathbf{r}_{j}dx_{i}$
as in (\ref{eq:inf_coord_tr}) and (\ref{eq:inf_coord_rot}). (We
do not distinguish upper and lower indices in this section.) Let $\mathit{vol}$
be the standard volume form in Euclidean space, then the $2$-forms
$A_{i}=\iota_{\partial_{i}}\mathrm{\mathit{vol}}$ form a basis of
$2$-forms satisfying $dx_{i}\wedge A_{j}=\delta_{ij}\mathrm{\mathrm{vol}}$,
moreover, $dA_{i}=0$ for Cartesian coordinates. Let $\left\{ \mathbf{v}_{i}^{*},\mathbf{r}_{i}^{*}\right\} $
be the dual basis of $\mathfrak{g}^{*}$ to $\left\{ \mathbf{v}_{i},\mathbf{r}_{i}\right\} $,
and expand $\Sigma$ as:

\begin{equation}
\Sigma=\sigma_{ij}\mathbf{v}_{j}^{*}A_{i}+\chi_{ij}\mathbf{r}_{j}^{*}A_{i}\label{eq:stress_local}
\end{equation}
where $\sigma_{ij}$ is the generalization of the Cauchy stress and
$\chi_{ij}$ represents moment stresses, the first index corresponds
to the area element while the second to the direction of force/moment
respectively. We have:

\begin{gather}
\langle e,\Sigma\rangle=\langle\varepsilon_{ij}\mathbf{v}_{j}dx_{i}+\tau_{ij}\mathbf{r}_{j}dx_{i},\sigma_{kl}\mathbf{v}_{l}^{*}A_{k}+\chi_{kl}\mathbf{r}_{l}^{*}A_{k}\rangle=\nonumber \\
=\left(\varepsilon_{ij}\sigma_{kl}\langle\mathbf{v}_{j},\mathbf{v}_{l}^{*}\rangle+\tau_{ij}\chi_{kl}\langle\mathbf{r}_{j},\mathbf{r}_{l}^{*}\rangle\right)dx_{i}\wedge A_{k}=\nonumber \\
\left(\varepsilon_{ij}\sigma_{kl}+\tau_{ij}\chi_{kl}\right)\delta_{jl}\delta_{ik}\mathrm{\mathit{vol}}=\left(\varepsilon_{ij}\sigma_{ij}+\tau_{ij}\chi_{ij}\right)\mathrm{\mathit{vol}}\label{eq:pairing_compute}
\end{gather}

Another important mathematical tool is the exterior covariant derivative
operator $D^{*}$, induced on $\tilde{W}^{*}$-valued forms on $\tilde{\mathcal{B}}$
by $\omega$, which will allow us to integrate by parts and write
down local forms of balance laws. There are multiple ways to define
it, one of them is through the Leibniz rule: for a $\tilde{W}^{*}$-valued
$p$-form $\Pi$, $D^{*}\Pi$ is the unique $\tilde{W}^{*}$-valued
$p+1$-form such that for any section $\xi$ of $\tilde{W}$ we have:
\citep{Kanso2007}

\begin{equation}
d\langle\xi,\Pi\rangle=\langle\xi,D^{*}\Pi\rangle+\langle D\xi,\Pi\rangle\label{eq:ext_cov_leib}
\end{equation}
This is very similar to how one induces the covariant derivative on
the cotangent bundle of a manifold from a linear connection on its
tangent bundle. Another possible definition comes from the observation
that $\tilde{W}^{*}$ is also an associated bundle to $\psi(\mathcal{P})$
by the coadjoint representation of $H$ on $\mathfrak{g}^{*}$. We
can again view any $\tilde{W}^{*}$-valued $p$-form $\Pi$ as a $\mathfrak{g}^{*}$-valued
equivariant horizontal $p$-form on $\psi(\mathcal{P})$, then its
exterior covariant derivative $D^{*}\Pi$ is nothing else but

\begin{equation}
D^{*}\Pi=d\Pi+\mathrm{ad}_{\omega}^{*}\Pi\label{eq:ext_deriv_coadj}
\end{equation}
where $\mathrm{ad}^{*}$ denotes the coadjoint representation of $\mathfrak{g}$
on $\mathfrak{g}^{*}$, defined via:

\begin{equation}
\langle\mathrm{ad}_{X}^{*}\mu,Y\rangle=-\langle\mu,\mathrm{ad}_{X}Y\rangle\quad\forall X,Y\in\mathfrak{g},\mu\in\mathfrak{g}^{*}\label{eq:coadj_def}
\end{equation}
The differential operator in (\ref{eq:ext_deriv_coadj}) is similar
to expressions which appear in the theory of the Euler-Poincar� equations
arising from the variation of group-invariant Lagrangians on Lie group
configuration spaces \citep{Carre2023,Marsden1999}.

Suppose that external volume forces and torques act on the Cosserat
solid, given by an $\tilde{W}^{*}$-valued $3$-form $F$ on the current
configuration $\tilde{\mathcal{B}}$, as well as traction forces and
couples on the boundary $\partial\tilde{\mathcal{B}}$, represented
by a $\tilde{W}^{*}$-valued $2$-form $T$. Let us impart a virtual
deformation $\xi$ on the current configuration $\tilde{\mathcal{B}}$,
given by a section of $\tilde{W}$, then the principle of virtual
work states that the total virtual work done by all forces and stresses
vanishes \citep{Germain1973,Germain2020,Eugster2017}. The three contributions
(neglecting inertia) to the virtual work are:
\begin{description}
\item [{a)}] The virtual work of external volume forces and torques: $\delta W_{ext}=\int_{\tilde{\mathcal{B}}}\langle\xi,F\rangle$.
\item [{b)}] The virtual work of traction forces and torques: $\delta W_{trac}=\int_{\partial\tilde{\mathcal{B}}}\langle\xi,T\rangle$.
\item [{c)}] The virtual work of stress in the bulk (the minus sign is
conventional): $\delta W_{int}=-\int_{\tilde{\mathcal{B}}}\langle D\xi,\Sigma\rangle$.
\end{description}
The principle of virtual work asserts that:

\begin{equation}
\delta W_{tot}=\delta W_{ext}+\delta W_{trac}+\delta W_{int}=0\label{eq:virt_work}
\end{equation}
for all virtual displacements $\xi$. Substituting the integral expressions
into (\ref{eq:virt_work}), using Stokes' theorem \citep{Frankel2011}
and the exterior covariant derivative (\ref{eq:ext_cov_leib}) to
integrate by parts yields:

\begin{gather}
\int_{\tilde{\mathcal{B}}}\langle\xi,F\rangle+\int_{\partial\tilde{\mathcal{B}}}\langle\xi,T\rangle-\int_{\mathcal{B}}\langle D\xi,\Sigma\rangle=\nonumber \\
=\int_{\tilde{\mathcal{B}}}\langle\xi,F+D^{*}\Sigma\rangle+\int_{\partial\tilde{\mathcal{B}}}\langle\xi,T-\Sigma\rangle=0\label{eq:virt_work_int}
\end{gather}
As (\ref{eq:virt_work_int}) holds for any virtual displacement field
$\xi$, we deduce the following equilibrium equations and boundary
conditions \citep{Schaefer1967}:

\begin{alignat}{1}
D^{*}\Sigma+F & =0\quad\text{on }\mathcal{\tilde{B}}\label{eq:equil_sp}\\
T & =\Sigma\quad\text{on }\partial\tilde{\mathcal{B}}\label{eq:equil_bc}
\end{alignat}
Equation (\ref{eq:equil_sp}) provides yet another interpretation
of the operator $D^{*}$: it is the generalization of the divergence
operator appearing in the usual Cauchy momentum equation. (\ref{eq:equil_sp})
is written in the Eulerian picture: to obtain the analougous Lagrangian
equilibrium equations, one can pull back (\ref{eq:equil_sp}) to the
reference configuration via $\psi$, the $W^{*}$-valued $2$-form
$S=\psi^{*}\Sigma$ is going to be the analogue of the second Piola-Kirchhoff
stress tensor in finite elasticity \citep{Marsden1994}. The first
Piola-Kirchhoff stress can be found from $\Sigma$ by pulling back
the form ``part'' from $\tilde{\mathcal{B}}$ to the reference configuration
$\mathcal{B}$ but doing nothing to the vector part \citep{Kanso2007}
(it is useful because it helps performing the integral (\ref{eq:virt_work_int})
in Lagrangian coordinates on $\mathcal{B}$).

For an explicit coordinate representation of (\ref{eq:equil_sp}),
let us work again in Cartesian coordinates, writing $F=\left(f_{i}\mathbf{v}_{i}^{*}+m_{i}\mathbf{r}_{i}^{*}\right)\mathit{vol}$
for the volume forces and torques. As in (\ref{eq:inf_comp}), $\omega$
is locally given by the $1$-form $\mathbf{v}_{i}dx_{i}$, thus using
(\ref{eq:stress_local}) and (\ref{eq:ext_deriv_coadj})

\begin{equation}
D^{*}\Sigma=d\left(\left(\sigma_{ij}\mathbf{v}_{j}^{*}+\chi_{ij}\mathbf{r}_{j}^{*}\right)A_{i}\right)+\mathrm{ad}_{\mathbf{v}_{k}dx_{k}}^{*}\left(\left(\sigma_{ij}\mathbf{v}_{j}^{*}+\chi_{ij}\mathbf{r}_{j}^{*}\right)A_{i}\right)\label{eq:stress_div_1}
\end{equation}
Using the commutation relations of $\mathfrak{g}$ we get $\mathrm{ad}_{\mathbf{v}_{i}}^{*}\mathbf{v}_{j}^{*}=\epsilon_{ijk}\mathbf{r}_{k}^{*}$
and $\mathrm{ad}_{\mathbf{v}_{i}}^{*}\mathbf{r}_{j}^{*}=0$, therefore:

\begin{equation}
D^{*}\Sigma=\left(\mathbf{v}_{j}^{*}\partial_{k}\sigma_{ij}+\mathbf{r}_{j}^{*}\partial_{k}\chi_{ij}+\mathbf{r}_{l}^{*}\epsilon_{kjl}\sigma_{ij}\right)dx_{k}\wedge A_{i}\label{eq:stress_div_2}
\end{equation}
Separating the coefficients of $\mathbf{v}_{i}^{*}$ and $\mathbf{r}_{j}^{*}$
and substituting into (\ref{eq:equil_sp}) we obtain the familiar
local balance of linear and angular momentum:

\begin{alignat}{1}
\partial_{j}\sigma_{ji}+f_{i} & =0\label{eq:force_balance}\\
\partial_{j}\chi_{ji}+\epsilon_{ijk}\sigma_{jk}+m_{i} & =0\label{eq:moment_balance}
\end{alignat}
When $F=0$, (\ref{eq:equil_sp}) takes the simple form $D^{*}\Sigma=0$.
Since $\omega$ is flat, $D^{*}D^{*}=0$ (just like in (\ref{eq:ext_cov_deriv_flat})),
so the solution is of the form $\Sigma=D^{*}Y$ for a stress potential
$Y$, which is a $\tilde{W}^{*}$-valued $1$-form that is only defined
up to a gauge transformation $Y\to Y+D^{*}\alpha$ for some section
$\alpha$ of $\tilde{W}^{*}$. While stress potentials for Cosserat
solids were derived in e.g. \citep{Cowin1970}, it is important to
note that similar potentials have been extensively utilized in many
areas of classical physics, such as scalar and vector potentials in
electromagnetism, Airy and Maxwell stress functions for certain problems
in Cauchy elasticity and the Papkovich-Neuber representiation of Stokes
flows. By formulating the theory of Cosserat solids in terms of these
stress potentials (and with the inclusion of inertia too), one obtains
a gauge theory similar to fracton theory in condensed matter physics
\citep{Gromov2020}. Equations (\ref{eq:equil_sp}) and (\ref{eq:equil_bc})
also suggest geometric structure-preserving numerical schemes using
discrete exterior calculus \citep{Hirani2003,Yavari2008,Boom2022}.

\section{Constitutive Relations\label{sec:Constitutive-Relations}}

In order to be able to close the above system of equations, one needs
to specify a constitutive law between stress and strain, characteristic
of the material in question \citep{Truesdell2004}. In general, these
can be quite complicated nonlinear relations which may depend on time
rates of tensors too. In this paper, we restrict our attention to
the simple case when the constitutive law gives the second Piola-Kirchhoff
stress $S$ at a point $X\in\mathcal{B}$ as a function $S(X,E(X))$
of $X$ and the strain measure $E$ at the point $X$. An important
class of materials is the set of hyperelastic materials, for which
the stress derives from a potential or stored energy function per
unit mass $U(X,E(X))$ as $S=\frac{\partial U}{\partial E}$. These
bodies are conservative: they do not perform any work along a closed
cycle in deformation space \citep{Marsden1994}.

The number of independent parameters in the constitutive law is restricted
by the symmetries of the material. In our setting, a material symmetry
is an $H$-bundle automorphism of $\mathcal{P}$ that leaves the functional
form of the constitutive relation invariant. More specificially, the
material symmetry group for a hyperelastic Cosserat solid at a point
$X\in\mathcal{B}$ is a subgroup $K_{X}\leq H$ of the orthogonal
group such that:

\begin{equation}
U(X,E(X))=U(X,L_{R}^{*}E(X))\quad\forall R\in K_{X}\label{eq:sym_group}
\end{equation}
for any possible strain $E(X)$, where $L_{R}:\mathcal{P}\to G$ is
a rigid rotation (and perhaps reflection) of the reference configuration
$\mathcal{P}$ by $R$ about $X$, i.e.:

\begin{equation}
L_{R}\left(\begin{bmatrix}1 & 0\\
x & S
\end{bmatrix}\right)=\begin{bmatrix}1 & 0\\
R(x-X) & RS
\end{bmatrix}\label{eq:mat_sym}
\end{equation}
The motivation behind definition (\ref{eq:sym_group}) is that the
action of a material symmetry is given by replacing $\psi$ with $\psi_{R}=\psi\circ L_{R}$
and the connection $\omega$ on $\mathcal{P}$ by $L_{R}^{*}\omega$
(c.f. rotating or reflecting the reference configuration); this way
$E$ changes to $E_{R}=\psi_{R}^{*}\omega-L_{R}^{*}\omega=L_{R}^{*}E$.

In components, $E$ is $\left(Q^{T}dy-dx,Q^{T}dQ\right)=\left(Ydx,\Gamma dx\right)$
from (\ref{eq:finite_strain_coords}), where $\Gamma$ corresponds
to the axial vector-valued wryness tensor, transforms to $E_{R}=\left(R^{T}YR,\left(\det R\right)R^{T}\Gamma R\right)$,
taking into account the axial vector nature of $\Gamma$ \citep{Eremeyev2012}.
Hence if the Cosserat solid is centrosymmetric (equivalent to achirality
in three dimensions), meaning $-I\in K_{X}$, then $U(X,E)=U(X,Y,\Gamma)=U(X,Y,-\Gamma)$,
so there cannot be any translation-rotation coupling in the constitutive
relation. If $K_{X}$ is a subgroup of $\mathrm{SO}(3)$, then the
material is not symmetric under reflections \citep{Lakes1982} and
that the microstructural constitutents are chiral. Other important
cases are when the material is isotropic so that $K_{X}=\mathrm{O(3)}$
or hemitropic when $K_{X}=\mathrm{SO}(3)$, in these cases the constitutive
relation can only involve certain invariants of the strain $E(X)$.
Chirality is one of the most exciting features of the theory of Cosserat
solids because it is ubiquitous in nature and soft matter \citep{Tan2022}
while entirely absent in Cauchy elasticity, and experimentally engineered
mechanical metamaterials have recently been demonstrated to exhibit
fascinating chiral effects such as acoustical activity \citep{Frenzel2017,Chen2020,Reasa2020},
confirming predictions of micropolar elasticity.

Linear constitutive relations are often a good approximation to constitutive
behaviour, especially when deformations are small. Suppose that in
local coordinates and with respect to a basis $\left\{ \mathbf{w}_{A}\right\} $
of $\mathfrak{g}$, we expand $E$ as a $\mathfrak{g}$-valued $1$-form
$E=E_{Ai}\mathbf{w}_{A}dx^{i}$. Using the dual basis $\left\{ \mathbf{w}_{A}^{*}\right\} $
of $\mathfrak{g}^{*}$ and a dual basis of $2$-forms $A_{i}$ as
in the previous section, we expand $S$ as the $\mathfrak{g}^{*}$-valued
$2$-form $S=S^{Bj}\mathbf{w}_{B}^{*}A_{j}$. A linear constitutive
law is then given by:

\begin{equation}
S^{Ai}=C^{ABij}E_{Bj}\label{eq:lin_const}
\end{equation}
for a general stiffness tensor $C^{ABij}$. Compared with the two
Lam� constants of classical elasticity, a linear isotropic or hemitropic
constitutive relation contains six or nine material constants respectively
\citep{Lakes1982}. If the material is hyperelastic, then $S^{Ai}=\partial U/\partial E_{Ai}$
for a stored energy function $U(X,E)$, and the symmetry of second
partial derivatives implies the major symmetry:

\begin{equation}
C^{ABij}=C^{BAji}\label{eq:major_sym}
\end{equation}
However, there are recent so-called odd elastic models of active solids
\citep{Scheibner2020} with rich phenomenology where the major symmetry
(\ref{eq:major_sym}) is broken and the material is able to perform
work along a closed cycle in deformation space via using energy produced
from an internal mechanism. Odd elasticity has also been extended
to and experimentally realized in micropolar beams with piezoelectric
activity \citep{Chen2021}. It is quite likely that most oriented
solids in nature \citep{Tan2022} also violate the symmetry (\ref{eq:major_sym}),
therefore it is an interesting future direction of research to study
further consequences of such an ``odd'' constitutive relation \citep{Surowka2022}.
Finally, for a more comprehensive discussion of constitutive relations
involving possible material symmetry groups and questions of material
uniformity and homogeneity in Cosserat media, we refer to e.g. \citep{Eremeyev2012,Epstein1998}.

\section{Discussion\label{sec:Discussion}}

In this paper a geometric theory of Cosserat solids has been outlined,
modelling the body as a principal fibre bundle $\mathcal{P}$ over
a three-dimensional base and $\mathrm{O(3)}$ as the typical fibre.
Configurations of the continuum were defined as bundle maps $\psi$
from $\mathcal{P}$ to an ambient $\mathrm{O(3)}$-bundle $G$, while
the strain measure as the difference between two Cartan connections:
a material one and the pullback of the Maurer-Cartan form on $G$
along $\psi$. Compatibility conditions were expressed using exterior
calculus, and incompatibilities were identified as the curvature of
the material connection on the Cartan geometry $\mathcal{P}$. Stress
was introduced as a dual quantity of strain, and balance laws were
obtained using the principle of virtual work. Constitutive laws relating
stress and strain were also briefly discussed, along with issues of
chirality and hyperelasticity.

Our work can be extended in a multitude of different directions. One
can take any Klein geometry $G/H$ and consider a generalized Cosserat
solid as a Cartan geometry modelled on $G/H$ (i.e. a principal $H$-bundle
$\mathcal{P}$ with a Cartan connection $\eta$) and bundle maps $\psi:\mathcal{P}\to G$
as configurations. (The ambient space can in principle be another
arbitrary Cartan geometry modelled on $G/H$). The notions of strain,
stress, compatibility conditions and constitutive relations can be
defined exactly analogously as it was done in the main text. For example,
micromorphic elasticity \citep{Mindlin1964,Eringen1999} can also
be cast in this language by taking $G$ to be the general affine group
$\mathrm{GA}(3)$ of line-preserving transformations of $\mathbb{E}^{3}$
and $H=\mathrm{GL}(3)$ the general linear group. However, the physical
relevance of such models is unclear as the introduction of the strain
measure was postulated based on invariance under rigid body transformations,
not under e.g. general affine transformations, not to mention the
potentially extremely large number of material parameters involved
in the constitutive relations. Nevertheless, one can certainly consider
two-dimensional Cosserat continua on the plane with $G=\mathrm{E(2)}$
and $H=\mathrm{O}(2)$ or on the $2$-sphere $S^{2}$ with $G=\mathrm{O}(3)$
and $H=\mathrm{O}(2)$. It is also possible to study other complex
materials using fibre bundles \citep{Capriz1989,Segev1994}, for example
configurations of polar liquid crystals can be viewed as bundle maps
between principal bundles over $\mathbb{E}^{3}$ and typical fibre
$S^{2}$. Nevertheless, these bundles are no longer principal, therefore
strain and stress measures become more involved for these media.

An obvious limitation of our theory is that it only considers static
deformations. A straightforward way to incorporate time is to make
the configuration $\psi$ time-dependent, then material velocity can
be defined as a vector field $V$ obtained by taking the partial derivative
of $\psi$ with respect to time. (An alternative and conceptually
cleaner approach would be to work in a covariant spacetime setting
\citep{Romano2014}). Analogously to (\ref{eq:inf_def_vec}) and (\ref{eq:inf_def_lie_alg}),
the velocity vector field $V$ can again be interpreted as a section
of a vector bundle over $\mathcal{B}$ with typical fibre $\mathfrak{g}$.
For the dynamical equations of motion one has to add an inertial contribution
to the virtual work principle (\ref{eq:virt_work}), usually deriving
from the variation of a kinetic energy, modelled as a bundle metric
on the vector bundle of velocities. However, the choice of this kinetic
energy metric is far from obvious \citep{Cowin1975,Cowin1980}: it
is usually assumed (by analogy with a standard rigid body) that it
is given by a mass density times the spatial metric $\rho g_{ij}$
on the translational part and $\rho I_{ij}$ on the rotational part
where $I_{ij}$ is a symmetric moment of inertia density tensor. In
general, one cannot rule out a term coupling angular and translational
velocites \citep{Eremeyev2020} or a more complicated expression for
the kinetic energy \citep{Capriz2003}. An additional issue is that
the evolution of inertial quantites (mass and moment of inertia density)
are usually governed by conservation laws, and while mass conservation
is a natural assumption, moment of inertia conservation as proposed
by \citet{Eringen1999} is more controversial \citep{Dahler1963}
(in particular, it has been suggested in the literature \citep{Vilchevskaya2019}
that one should consider moment of inertia production terms as well).
In any case, most experimental systems of interest only undergo small
deformations or are overdamped, in which case either a linearised
treatment -- where the inertial terms can be added in straightforwardly
-- is sufficient, or inertia terms are absent entirely and one may
include viscous damping terms on a phenomenological basis \citep{Lakes2001}.

We conclude by mentioning that the geometrisation of Schaefer's theory
of the Cosserat solid immediately suggests methods for numerically
integration that preserve geometric structures \citep{Hirani2003,Yavari2008,Boom2022}.
We shall address this in forthcoming work where the geometrisation
will be implemented in the alternative setting of a field theory \citep{Kikuchi2022}. 

\begin{acknowledgements}
	We thank Professor M. E. Cates and Lukas Kikuchi for many helpful discussions and a critical reading of the manuscript.
	This work was supported by the Engineering and Physical Sciences Research Council (UK) through a studentship for the first author.
\end{acknowledgements}

\bibliographystyle{apsrev4-2}
\bibliography{article_pre}

\section*{Appendix A: Informal Motivation of Strain Measures\label{sec:AppendixA}}

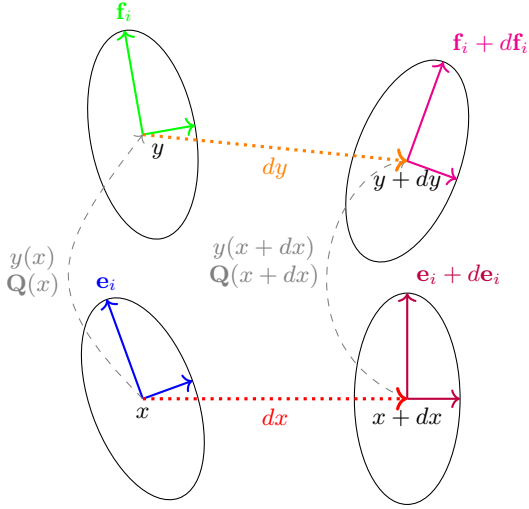
\begin{figure}
\begin{tikzpicture}
	\coordinate[label=below:{$x$}](x) at (0,0);
	\coordinate[label=below:{$x+dx$}](z) at (100pt,0);
	\coordinate[label=below right:{$y$}](y) at (0,100pt);
	\coordinate[label=below:{$y+dy$}](w) at (100pt,90pt);
	
	\draw[rotate=20] (x) ellipse [x radius=20pt, y radius=40pt];
	\draw[->,blue,thick,rotate=20] (0,0)--(0,40pt) node[above] {$\mathbf{e}_i$};
	\draw[->,blue,thick,rotate=20] (0,0)--(20pt,0);
	
	\draw (z) ellipse [x radius=20pt, y radius=40pt];
	\draw[->,purple,thick] (100pt,0)--(100pt,40pt) node[above right] {$\mathbf{e}_i+d\mathbf{e}_i$};
	\draw[->,purple,thick] (100pt,0)--(120pt,0);
	
	\draw (y) [rotate around={10:(y)}] ellipse [x radius=20pt, y radius=40pt];
	\draw[->,green,thick,rotate around={10:(y)}] (0,100pt)--(0,140pt) node[above] {$\mathbf{f}_i$};
	\draw[->,green,thick,rotate around={10:(y)}] (0,100pt)--(20pt,100pt);
	
	\draw (w) [rotate around={-20:(w)}] ellipse [x radius=20pt, y radius=40pt];
	\draw[->,magenta,thick,rotate around={-20:(w)}] (100pt,90pt)--(100pt,130pt) node[above right] {$\mathbf{f}_i+d\mathbf{f}_i$};
	\draw[->,magenta,thick,rotate around={-20:(w)}] (100pt,90pt)--(120pt,90pt);

	\draw[->,gray,dashed] (x) .. controls (-40pt,40pt) and (-35pt,55pt) .. node[left,align=center] {$y(x)$\\$\mathbf{Q}(x)$} (y);
	\draw[->,gray,dashed] (z) .. controls (50pt,20pt) and (70pt,88pt) .. node[left, align=center] {$y(x+dx)$\\$\mathbf{Q}(x+dx)$} (w);
	\draw[->,very thick,red,dotted] (x)-- node[below] {$dx$} (z);
	\draw[->,very thick,orange,dotted] (y)-- node[below] {$dy$} (w);
\end{tikzpicture}\caption{Illustration of deformation in Cosserat medium. The translational
strain can be measured by looking at the change in the angle between
the vectors $\mathbf{e}_{i}$ and $dx$ as $\mathbf{e}_{i}$ goes
to $\mathbf{f}_{i}$ and $dx$ goes to $dy$. Similarly, rotational
strain is captured by the change in the relative angle of infinitesimally
close directors $\mathbf{e}_{i}$ and $\mathbf{e}_{i}+d\mathbf{e}_{i}$
as they go to $\mathbf{f}_{i}$ and $\mathbf{f}_{i}+d\mathbf{f}_{i}$
respectively.}
\end{figure}
In this appedix we give a heuristic derivation of the infinitesimal
strain measures, based on Cartan's method of moving frames \citep{Cartan1923}.
This rationalizes as to why the strain measures can be considered
as differences of connections. Here we view the Cosserat body as a
continuum whose material points have an orthonormal set of directors
attached to them, in line with the classical treatment of \citet{Ericksen1957}
and others. Consider two infinitesimally close points $x$ and $x+dx$
of a Cosserat solid, with corresponding directors $\mathbf{e}_{i}$
and $\mathbf{e}_{i}+d\mathbf{e}_{i}$. Now suppose that the body undergoes
an infinitesimal deformation, such that $x$ gets mapped to $y$,
$x+dx$ gets mapped to $y+dy$, and the directors transform to $\mathbf{f}_{i}$
and $\mathbf{f}_{i}+d\mathbf{f}_{i}$. In the micropolar theory, the
directors are assumed to be rigid, which means that $\mathbf{f}_{i}=\mathbf{Q}(\mathbf{e}_{i})$
for some proper orthogonal tensor $\mathbf{Q}$. If we view $dx$
and $dy$ as infinitesimal vectors based at $x$ and $y$ respectively,
we can write:

\begin{alignat}{1}
dx & =\mathbf{e}_{i}\theta^{i}\label{eq:solder1}\\
dy & =\mathbf{f}_{i}\widetilde{\theta}^{i}\label{eq:solder2}
\end{alignat}
Similarly, for the infinitesimal relative orientations of the directors
we get:
\begin{alignat}{1}
d\mathbf{e}_{i} & =\mathbf{e}_{j}\Omega_{i}^{j}\label{eq:conn1}\\
d\mathbf{f}_{i} & =\mathbf{f}_{j}\widetilde{\Omega}_{i}^{j}\label{eq:conn2}
\end{alignat}
The deformation is encoded in the relative changes $\Delta\theta^{i}=\widetilde{\theta}^{i}-\theta^{i}$
and $\Delta\Omega_{i}^{j}=\widetilde{\Omega}_{i}^{j}-\Omega_{i}^{j}$.
We cannot directly subtract $dy$ and $dx$ or $d\mathbf{f}_{i}$
and $d\mathbf{e}_{i}$ because they live in different vector spaces.
However, we can use $\mathbf{Q}^{-1}=\mathbf{Q}^{T}$ to bring back
the directors $\mathbf{f}_{i}$ to $x$. This way we find the following
deformation measures:

\begin{alignat}{1}
\mathbf{Q}^{T}dy-dx & =\mathbf{Q}^{T}\left(\mathbf{f}_{i}\widetilde{\theta}^{i}\right)-\mathbf{e}_{i}\theta^{i}=\mathbf{e}_{i}\left(\Delta\theta^{i}\right)\label{eq:diff_sold}\\
\mathbf{Q}^{T}(d\mathbf{f}_{i})-d\mathbf{e}_{i} & =\mathbf{Q}^{T}\left(\mathbf{f}_{j}\widetilde{\Omega}_{i}^{j}\right)-\mathbf{e}_{j}\Omega_{i}^{j}=\mathbf{e}_{j}\left(\Delta\Omega_{i}^{j}\right)\label{eq:diff_conn}
\end{alignat}
(\ref{eq:diff_conn}) can also be written as:
\begin{alignat*}{1}
\mathbf{Q}^{T}(d(\mathbf{Q}(\mathbf{e}_{i}))-d\mathbf{e}_{i}=\mathbf{Q}^{T}d\mathbf{Q}(\mathbf{e}_{i})
\end{alignat*}
Hence we recover the usual strain measures $\mathbf{Q}^{T}dy-dx$
and $\mathbf{Q}^{T}d\mathbf{Q}$ \citep{Pietraszkiewicz2009,Altenbach2013}.
Now suppose that $\mathbf{Q}=\mathbf{I}+t\mathbf{\hat{\omega}}+\mathcal{O}(t^{2})$
and $y=x+t\mathbf{v}+\mathcal{O}(t^{2})$, where $\hat{\omega}\in\mathfrak{so}(3)$
corresponding to $\omega\in\mathbb{R}^{3}$ then we find:

\begin{alignat*}{1}
\mathbf{Q}^{T}dy-dx & =(\mathbf{I}-t\mathbf{\hat{\omega}})(dx+t\mathbf{v})-dx\\
 & =t(\mathbf{v}+dx\times\omega)+\mathcal{O}\left(t^{2}\right)\\
\mathbf{Q}^{T}d\mathbf{Q} & =(\mathbf{I}-t\mathbf{\hat{\omega}})t(d\hat{\omega})+\mathcal{O}(t^{2})=td\hat{\omega}+\mathcal{O}\left(t^{2}\right)
\end{alignat*}
The infinitesimal strain measures are recovered if the derivative
wrt. $t$ is taken as $t\to0$.

We can also phrase the above discussion as follows. Let the tuple
$E=(x,\mathbf{e}_{i})$ be a frame at $x$, viewed as a row vector
formed from $x$ and $\mathbf{e}_{i}$, similarly $F=(y,\mathbf{f}_{i})$.
Then we have:

\begin{alignat}{1}
dE & =(dx,d\mathbf{e}_{j})=(x,\mathbf{e}_{j})\cdot\begin{pmatrix}0 & 0\\
\theta^{j} & \Omega_{i}^{j}
\end{pmatrix}\triangleq E\cdot\xi\label{eq:cart_conn_1}\\
dF & =(dy,d\mathbf{f}_{j})=(y,\mathbf{f}_{j})\cdot\begin{pmatrix}0 & 0\\
\widetilde{\theta}^{j} & \widetilde{\Omega}_{i}^{j}
\end{pmatrix}\triangleq F\cdot\widetilde{\xi}\label{eq:cart_conn_2}
\end{alignat}
Then our strain measures can be viewed together as:

\begin{equation}
\Gamma=(x,\mathbf{e}_{i})\cdot\left(\widetilde{\xi}-\xi\right)=E\cdot\left(F^{-1}dF-E^{-1}dE\right)\label{eq:cart_conn_diff}
\end{equation}

Equations (\ref{eq:solder1}),(\ref{eq:solder2}),(\ref{eq:conn1})
and (\ref{eq:conn2}) are known as Cartan's structure equations, describing
an affine connection with in a local trivialization (i.e. a choice
of directors $\mathbf{e}_{i}$ and $\mathbf{f}_{i}$) using vector-valued
differential forms \citep{Cartan1923,Frankel2011}. The deformation
measures are essentially differences of the material connection and
solder forms $\Omega_{i}^{j}$ and $\theta^{i}$ and the pullback
connection and solder forms $\widetilde{\Omega}_{i}^{j}$ and $\widetilde{\theta}^{i}$.
Equations (\ref{eq:cart_conn_1})--(\ref{eq:cart_conn_2}) also highlight
the fact that Cartan connections ``encapsulate'' the solder forms
and connection forms into a larger Lie algebra-valued form. 
\end{document}